\documentclass{aastex}
\usepackage{emulateapj5}

\begin{document}

\submitted{Accepted for publication in ApJ}
\title{Heating and Ionization of the Intergalactic Medium \\ by an Early
X-Ray Background} 
\author{Aparna Venkatesan, Mark L. Giroux\altaffilmark{1} \& J. Michael Shull} 
\affil{CASA, Department of Astrophysical and Planetary Sciences, \\
389 UCB, University of Colorado,  Boulder, CO 80309-0389}
\altaffiltext{1}{Now at Department of Physics and Astronomy, East Tennessee
State University.}
\email{aparna@casa.colorado.edu, giroux@casa.colorado.edu, mshull@casa.colorado.edu}
\vspace{0.1in}

\begin{abstract}

Observational studies indicate that the intergalactic medium (IGM) is
highly ionized up to redshifts just over 6. A number of models have been
developed to describe the process of reionization and the effects of the
ionizing photons from the first luminous objects. In this paper, we study
the impact of an X-ray background, such as high-energy photons from early
quasars, on the temperature and ionization of the IGM prior to
reionization, before the fully ionized bubbles associated with individual
sources have overlapped.  X-rays, which have large mean free paths relative
to EUV photons, and their photoelectrons can have significant effects on
the thermal and ionization balance. We find that hydrogen ionization is
dominated by the X-ray photoionization of neutral helium and the resulting
secondary electrons. Thus, the IGM may have been warm and weakly ionized
prior to full reionization. We examine several related consequences,
including the filtering of the baryonic Jeans mass scale, signatures in the
cosmic microwave background, and the H$^{-}$-catalyzed production of
molecular hydrogen.

\end{abstract}

\section{Introduction}

Observations of the spectra of high-redshift quasars and galaxies
shortward of Ly$\alpha$ have revealed that the IGM is highly ionized
up to redshifts, $z$ $\sim$ 6 \citep{fan1,hugal,spinrad}. The
reionization of the IGM subsequent to recombination at $z \sim$ 1000
is thought to have been caused by increasing numbers of the first
luminous sources, such as stars \citep{hl97,fuka}, quasars
\citep{donshull87,valsilk}, protogalaxies
\citep{gs96,madrees99,ciardi,chiuost,gn00}, or combinations of these
\citep{gs96,tsb94}.

It is unclear whether the H~I and He~II reionization epochs were coeval; it
is certainly possible with a sufficiently hard ionizing source spectrum,
from, e.g, quasars or zero-metallicity stars \citep{gs96,tumshull}. The
present data, however, indicate that He~II reionization occurred at $z \sim
3$ \citep{jakobsen,hog97,reim97,kriss}, and that of hydrogen before $z \sim
6$. The degree to which QSOs influence the process of reionization depends
on their relative abundance at high redshifts. There have been indications
that the QSO space density decreases beyond a peak at $z \sim 3$. This has
recently been corroborated by the Sloan Digital Sky Survey (SDSS)
observations of high-$z$ QSOs \citep{fan2, fan3}. However, QSOs may still
be relevant to reionization, if they are powered by massive black holes
that are postulated to form as a fixed universal fraction of the mass of
collapsing haloes at all redshifts (see, e.g., \citet{magor,korm}), leading
to a large population of mini-quasars at $z \ga 6$ \citep{hl98}. The
observed turnover in the QSO space density at $z \ga 3$ would then be true
only for the brightest QSOs.

It is also possible that some high-$z$ QSOs have been missed owing to dust
obscuration.  The most plausible scenario rests on evidence that the X-ray
background includes a substantial contribution from strongly absorbed AGN
with hard X-ray spectra \citep{gilli}.  Recent results from the {\it
Chandra} satellite \citep{mush, giac} also suggest that many of the
resolved hard X-ray sources in the background could be AGN in
dust-enshrouded galaxies or a population of quasars at extremely high
redshift. A contrary view is provided by \citet{shaver} who used a sample
of 442 radio-loud quasars, unhindered by dust selection effects, to confirm
the rapid decline in quasar space density at $z > 2.5$.

In this paper, we study the effects of an X-ray background (XRB), including
the case of X-ray photons generated by high-$z$ quasars, on the temperature
and ionization of the IGM prior to full reionization.  X-rays have much
larger mean free paths than extreme ultraviolet (EUV) photons with energies
$\geq$ 13.6 eV, and can permeate the IGM relatively uniformly. Thus, they
are capable of significantly heating the IGM prior to the epoch when the
bubbles that are highly ionized by EUV photons around individual sources
overlap. Owing both to the lower photoionization cross sections of H~I and
He~II and to the reduced intensity of the radiation field at X-ray,
relative to EUV energies, X-rays alone generally do not produce a fully
ionized IGM. However, the primary and secondary ionizations created by
X-rays \citep{svs85} can produce a mild increase in temperature and partial
reionization.

A number of authors have examined the reheating of the IGM in association
with or preceding reionization itself \citep{go97,valsilk,sgb94,gs96},
while others have studied the effects of X-rays on the high-$z$ IGM
\citep{cowsik,coll91,me99,har,oh}. The new points of interest presented in
this work include: self-consistent solutions for the IGM temperature and
ionization resulting purely from an XRB; the significant effects of
secondary electrons; and the use of more accurate calculations of
photoionization cross sections and recombination rate coefficients. We find
that the pre-reionization universe may be described by a model in which the
first luminous sources and their individual EUV Str\"{o}mgren spheres are
embedded in a pre-heated, partially ionized IGM, rather than in a cold,
completely neutral IGM. However, an early XRB does not create significant
amounts of new molecular hydrogen in the IGM, despite the increased
population of free electrons. This is primarily due to the strong
photo-destruction of H$^{-}$ and H$_2$ by the near IR/optical (henceforth
IR/O) and the far-UV (UV photons in the Lyman-Werner bands; henceforth FUV)
backgrounds, in the respective energy ranges 0.755--11.2 eV and 11.2--13.6
eV, associated with any XRB generated by QSOs. We also find that the
enhanced electron fraction in the IGM arising from an XRB prior to
reionization may cause an overestimation of the reionization epoch as
determined from the cosmic microwave background (CMB).

The plan of this paper is as follows. In \S 2, we outline the IGM
heating and cooling processes that we consider, and the models for
high-$z$ XRBs. In \S 3, we present our results and discuss their
implications for the cosmological Jeans mass, the CMB, the H$^{-}$
catalysis of molecular hydrogen, and the inhibition of galactic
outflows. We summarize our results in \S 4.

\section{X-Ray Photoionization Model}

In this section, we detail our assumptions for the XRB and ionizing QSO
spectra, and we describe the various processes used to determine
self-consistent IGM temperatures and ionization fractions of hydrogen and
helium. We focus on the heating and ionizing effects of the X-rays on the
initially neutral background IGM, but we do not explicitly follow the
growth of individual fully ionized regions around the first QSOs or
star-forming galaxies. As reionization begins, the volume filling factor of
such highly ionized regions is small. The remaining IGM is cold and
neutral, with only a small residual electron fraction ($\sim 10^{-4}$) from
the recombination epoch \citep{seager}.  Highly energetic photons emitted
by these sources will propagate much more quickly than their associated
ionization fronts.  This implies that an epoch will exist when the
thermodynamic properties of most of the IGM may be determined by these
X-rays.

We model the properties of this volume by solving for ionization and
thermal evolution in a uniform primordial medium.  We consider the
following processes for ionization solutions: photoionization,
collisional ionization, case B radiative recombination, dielectronic
recombination for He~I, and the coupling between H and He caused by
the radiation fields from the He~I 24.6 eV recombination continuum and
from the bound-bound transitions of He~I (photon energies at 19.8 eV,
21.2 eV, and the two-photon continuum with an energy sum of 20.6
eV). The last of these processes is incorporated following the method
in \citet{ost}. The exact form of the photoionization cross sections
for H~I and He~II are taken from \citet{spitzer}; we use the fit from
\citet{verner} for He~I.

We also account for the effects of the secondary ionizations and
excitations of H~I and He~I due to the energetic photoelectrons
liberated by the X-rays (\citet{svs85}; henceforth SVS85). These
introduce a further coupling between the ionization equilibria of H
and He.  Secondary ionization dominates over direct photoionization
for H~I in the specialized circumstance where X-rays are the sole
source of photoionization. A typical X-ray photon is far more likely
to be absorbed by He~I rather than H~I.  The ejected photoelectron,
however, will ionize many more H~I atoms than He~I, as H~I is more
abundant.  As a result, secondary ionizations from He~I photoelectrons
and the radiation associated with He~I recombination and excitation
are the primary sources of H~I ionization.  As pointed out by SVS85,
the partitioning of a primary electron's energy, between heating the
gas and secondary ionizations and excitations of H~I and He~I, is
itself a function of the gas ionization fraction, $x$. As $x$
decreases, and in particular for $x \la 0.1$, the photoelectron
deposits more of its energy in collisional ionizations/excitations and
less in heat. This $x$-dependence is therefore important to track in
the pre-reionization IGM, so that the role of secondary ionizations
for H~I is not underestimated. In order to be consistent with SVS85,
who assumed that the ionization fractions of hydrogen and once-ionized
helium were equal, appropriate for the interstellar environment they
were considering, we have recast the dependence in their results on
$x_{\rm H^+}$ to a direct dependence on the electron fraction, $x_{\rm
e}$ = $n_{\rm e}/(n_{\rm H} + n_{\rm He})$, where the number density
of electrons $n_{\rm e} = n_{\rm H II} + n_{\rm He II} + 2 n_{\rm He
III}$.

We compute the thermal evolution of the IGM including the following
processes: photoelectric heating from the secondary electrons of H and He,
as prescribed by SVS85 (the SVS85 solution generates less heating relative
to a prescription where 100\% of the photoelectron's excess energy goes
into heating the IGM), and heating from the H~I photoelectrons liberated by
the bound-bound transitions or the 24.6 eV recombination continuum of He I
(here the loss of excess energy to heat is taken to be 100\% as further
ionizations of H~I or He~I are not possible). Cooling terms include
radiative and dielectronic recombination, thermal bremsstrahlung, Compton
scattering off the CMB, collisional ionization and excitation, and the
adiabatic expansion of the IGM.  The values of the recombination and
cooling coefficients for temperatures $\la$ 10$^4$ K were taken from
\citet{hum94} and \citet{humst98}, and the heating contribution from the He
I two-photon process was calculated using the photon frequency distribution
given in \citet{drake69}. For the purposes of comparison, we will also
consider cases that do not include adiabatic cooling, which is mimicked by
artificially holding the IGM density constant at its value at $z = 10$, so
that there is no expansion cooling for $z < 10$. This would correspond to
regions of the IGM that have ceased to participate in the Hubble expansion,
but that have not yet begun to collapse.

The magnitude and spectral shape of the XRB at high redshift is unknown,
and a completely self-consistent model of its formation and evolution is
beyond the scope of this paper.  Although we suggest motivations for our
extrapolations of an XRB to early epochs, the three cases below merely
provide a theoretical framework in which to study the effects of an XRB at
high redshifts. We adopt two basic forms for the high-$z$ XRB, one that is
more directly tied to early QSOs, and the other, a modified version of the
XRB given in \citet{me99}.  In the first case, the specific intensity of
each QSO is taken to have a broken power-law form, $I_\nu \propto
\nu^{-\alpha}$, where $\alpha$ = 1.8 for $h \nu$ = 13.6--300 eV
\citep{zheng} and $\alpha$ = 0.8 for $h \nu$ = 300 eV - 10 keV
\citep{wash}. We assume that only those photons with energies $\ga$ 1 keV
permeate the IGM uniformly, while the less energetic photons are absorbed
in the individual host halos. Each QSO's spectrum is further attenuated by
absorption by hydrogen or helium in the IGM. If we take the comoving number
density of QSOs to have a peak abundance of $\Phi_{\rm QSO}$ = 10$^{-6}$
Mpc$^{-3}$ at $z = 3$ and $f_{\rm QSO} \Phi_{\rm QSO}$ at earlier epochs,
then the mean physical (not comoving) separation between QSOs at a given
$z$ is:

\begin{eqnarray}
d_{\rm QSO} = (100 \; {\rm Mpc}) \; f_{\rm QSO}^{-1/3} (1 + z)^{-1} .
\end{eqnarray}
\vspace{0.05in}

We assume then that the cumulative specific intensity in the IGM
evolves as $d_{\rm QSO}^2$, with the evolution of the source density
already factored in through $f_{\rm QSO}$. We normalize the unattenuated
mean intensity $I_\nu$ such that an extrapolation down to 13.6 eV,
using the QSO's intrinsic spectral shape, is $I_0$ = 10$^{-21}$ erg
cm$^{-2}$ s$^{-1}$ Hz$^{-1}$ sr$^{-1}$ at $z = 3$, consistent with
observational constraints (see, e.g., \citet{hm96,fard}, and
references therein). Realistically, this underestimates the
corresponding X-ray contribution because the radiation at 13.6 eV is
subject to some attenuation from the IGM, even at $z = 3$.

Combining the above factors, we may parametrize the evolution of the
IGM-filtered specific intensity for $z \geq 3$ as:

\begin{eqnarray}
I_\nu  & = & 3.8 \times 10^{-24} \, (h \nu/300 \, {\rm eV})^{-0.8} \; f_{\rm QSO}^{2/3} \; e^{-\tau_\nu} \nonumber \\ & & \times \; [(1 + z)/4]^2 \; {\rm erg \; cm^{-2} \; s^{-1} \; Hz^{-1} \; sr^{-1}},
\end{eqnarray}
\vspace{0.05in}

\noindent where we adopt $f_{\rm QSO}$ = $10^{- 0.5(z - 3)}$ from the
SDSS collaboration \citep{fan2}, and $\tau_\nu = d_{\rm QSO} [n_{\rm
H^0} \sigma_\nu ({\rm H^0}) + n_{\rm He^0} \sigma_\nu ({\rm He^0}) +
n_{\rm He^+} \sigma_\nu ({\rm He^+})]$ is the IGM optical depth, with
$\sigma_\nu$ being the appropriate photoionization cross section for
each species. Henceforth, we refer to the above XRB as case S (for the
Sloan survey).

An alternate prescription for the high-$z$ XRB is given in
\citet{me99}, in which the present-day XRB evolves as $(1 + z)^3$,
weighted by an exponential cutoff factor to account for the decreasing
source density at high redshift.  The specific intensity of this XRB
is given by eqn. [6] of \citet{me99} in units of keV cm$^{-2}$
s$^{-1}$ keV$^{-1}$ sr$^{-1}$, which, over the energy range 1--10 keV,
we take to be:

\begin{eqnarray}
I_\nu & = & 7.7 \; (h \nu/1 \, {\rm keV})^{-0.29} \; {\rm exp}(- h \nu/40 \, {\rm keV}) \nonumber \\ & & \times \; (1 + z)^{3} \; {\rm exp}[- (z/z_c)^2].
\end{eqnarray}
\vspace{0.05in}

We will consider two cases of this XRB: case 1 with $z_c$ = 5, as in
\citet{me99}, and case 2 without the exponential cutoff factor, as in
\citet{coll91} where the XRB evolves cosmologically as a radiation
field. The assumed specific intensity of each XRB increases from case
S through cases 1 and 2. At $ z = 9$, for example, their respective
values at 2 keV are 4.4 $\times 10^{-26}$, 1.6 $\times 10^{-24}$, and
4 $\times 10^{-23}$ erg cm$^{-2}$ s$^{-1}$ Hz$^{-1}$ sr$^{-1}$.

The evolution of the temperature and ionization fractions is calculated as
follows. We take the background cosmology to be described by, $\Omega_{\rm
m}$ = 0.3, $\Omega_\Lambda$ = 0.7, $h = 0.7$, $\Omega_{\rm b} h^2$ = 0.019,
$T_{\rm CMB,0}$ = 2.728 K, and $Y_{\rm He}$ = 0.24.  The calculation is
begun at $z=12$, with an initial IGM temperature of 20 K and residual
ionization fractions of $x_{\rm H^+} = 10^{-4}$ and $x_{\rm He^+} =
10^{-9}$. We consider XRBs in the redshift range 7 $< z <$ 12; the lower
limit is roughly the latest epoch from observational limits \citep{fan3,
becker} that can be considered as being prior to reionization, while at $z
= 12$ horizon effects begin to attenuate the XRB. We demonstrate the latter
by equating the light-crossing time for half the mean physical source
separation (given by eqn. 1) with the age of the universe for XRB case S,
accounting for the finite time needed for the X-ray photons from individual
sources to permeate the IGM.  As case S is the weakest XRB considered here,
with the lowest source density, this yields a conservative lower limit to
the redshift at which horizon effects become important. Furthermore, the
Eddington accretion timescale is $\sim$ 4.5 $\times 10^7$ yr, assuming an
average QSO radiative efficiency of 10\%. Since this corresponds to $z \sim
50$ in our adopted cosmology, by $z = 12$, sources have had $\sim$ 4
$\times 10^8$ yr (many times their Eddington timescale) to form and
generate an XRB (see, however, \citet{hl01}, on how this may not prove
sufficient for the formation of the most massive black holes that likely
power the brightest QSOs).

Our numerical solution for the evolving thermodynamic properties of the IGM
solves the nonequilibrium rate equations and energy equation for the
expanding IGM [cf. \citet{gs96}, eqns. (2.1)-(2.4)].  We do not follow the
nonequilibrium formation and destruction of H$_2$, and we do not consider
detailed radiative transfer.  The equations are integrated with a
fourth-order Runge-Kutta scheme with the time steps set to be less than 3\%
of the shortest relevant thermal or ionization timescale.

\section{Results and Implications}

In Figures 1--3, we show the temperature, and hydrogen and helium
ionization fractions of the IGM as a function of redshift for XRB cases 1
and 2. Case S was found to have negligible effects on the thermal and
ionization properties of the IGM, and did not alter them appreciably, over
the redshifts that we consider, from their values in an expanding IGM
without any X-rays. We therefore do not include case S in the figures, but
emphasize that our first result is that there will be no significant
effects on the evolving IGM for a conservative estimate of the
pre-reionization XRB.

The plots show that, depending on the choice of the XRB, the IGM
temperature ranges from $\sim$ 100 K to $\sim$ 10$^4$ K, with $x_{\rm
H^+}$ varying from 0.1\% to $\sim$ 20\%. For each case, the solutions
including and excluding adiabatic cooling are almost identical, and
are practically overlaid in the figures. As may be expected, the
stronger the XRB, the greater are the heating and ionization of the
IGM. Though we do not show it here, an XRB whose magnitude lies
between those of cases 1 and 2 generates IGM temperatures and
ionization fractions that are bracketed by the two cases shown in the
figures.

\centerline{\epsfxsize=1.0\hsize{\epsfbox{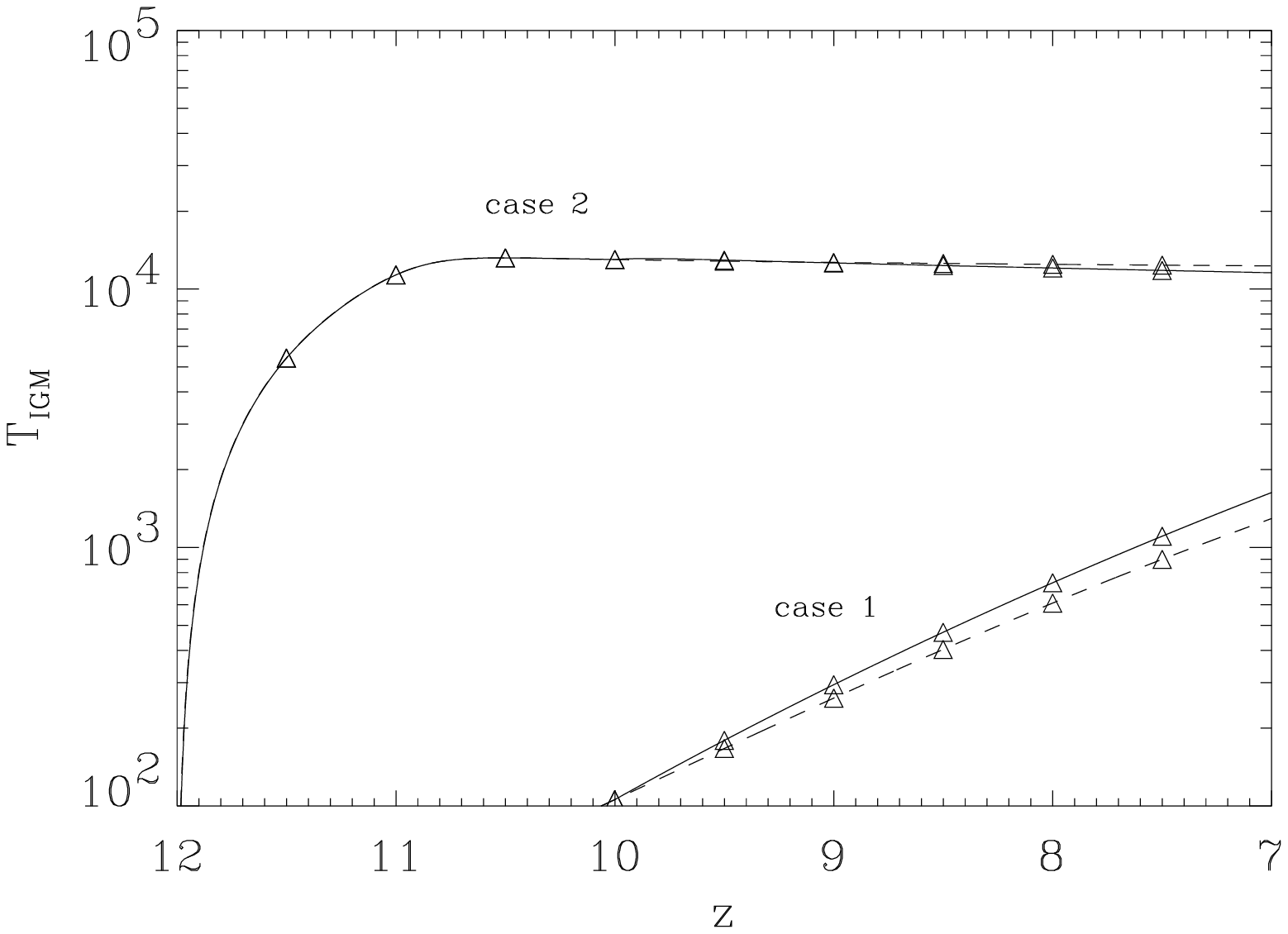}}}

\vspace{0.1in}
\figcaption{The IGM temperature is shown as a function of redshift for
the two XRBs described in the text, each with (dashed lines) and
without (solid lines) adiabatic cooling from cosmological expansion.}
\vspace{0.2in} 

\centerline{\epsfxsize=1.0\hsize{\epsfbox{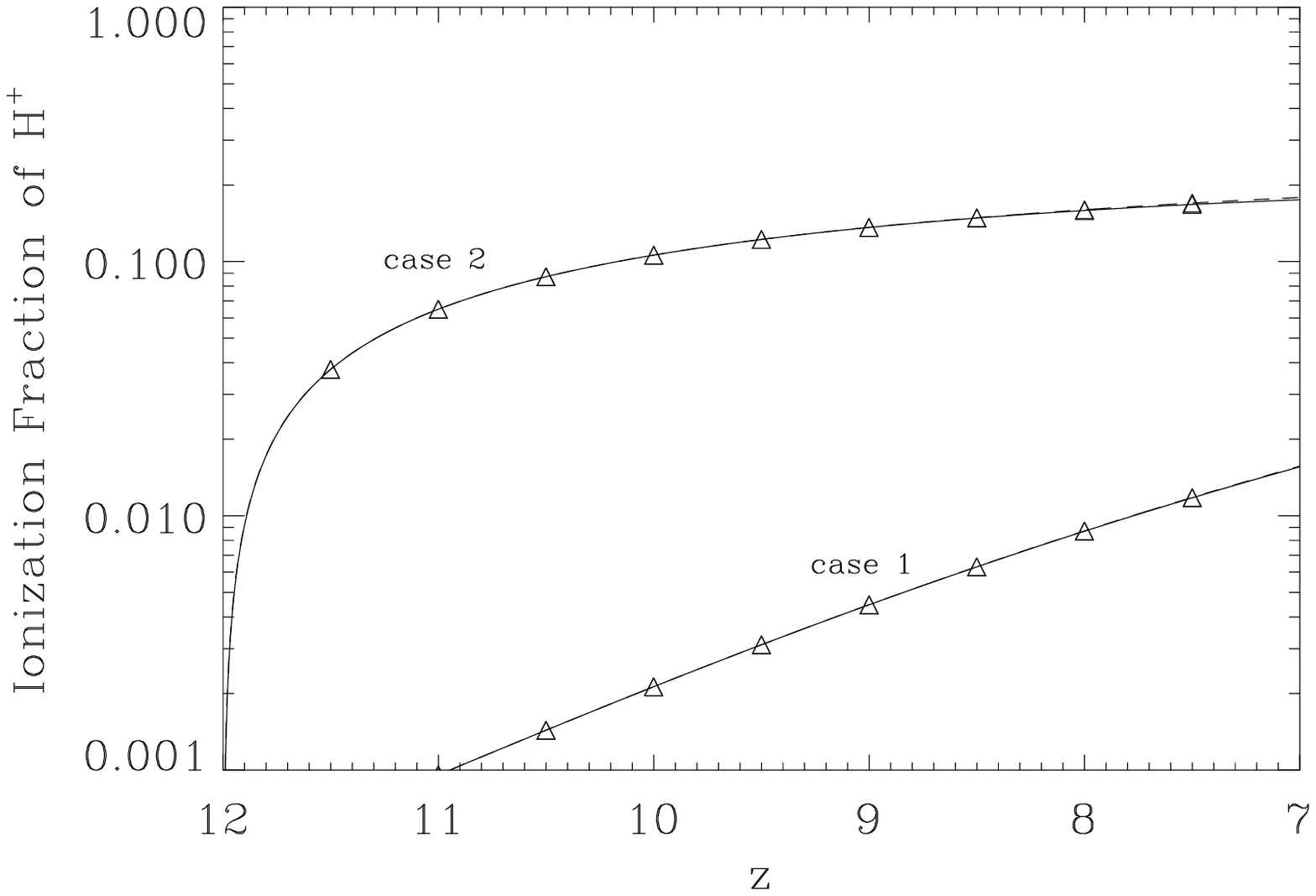}}}

\vspace{0.1in}
\figcaption{The IGM ionization fraction of hydrogen is shown as a
function of redshift for the two XRBs described in the text, each with
(dashed lines) and without (solid lines) adiabatic cooling from
cosmological expansion.}
\vspace{0.2in}

Note that for case 2, $T_{\rm IGM}$ is slightly higher when adiabatic
cooling is included rather than excluded. This counter-intuitive result
arises from our choice to mimic the cases without adiabatic cooling by
stopping the cosmological evolution of the IGM density at $z = 10$.  For
case 2, by $z = 10$, cooling is dominated by line excitation cooling, which
is proportional to the square of the IGM density, and not adiabatic
cooling.  Consequently, for the cases including adiabatic cooling, where
the IGM density continues to decrease at $z < 10$, the cooling decreases
more rapidly than it does in the absence of adiabatic cooling, leading to
marginally larger values of $T_{\rm IGM}$. Note also that when $x_{\rm e}$,
which roughly tracks $x_{\rm H^+}$, exceeds about 10\% (case 2), secondary
ionization becomes less important for H~I, and $x_{\rm He^+}$ begins to
exceed $x_{\rm H^+}$.  We find that the He~III fraction $x_{\rm He^{+2}}$
was typically smaller by a factor of 100 than $x_{\rm He^+}$, and we do not
display it in the figures.

\centerline{\epsfxsize=1.0\hsize{\epsfbox{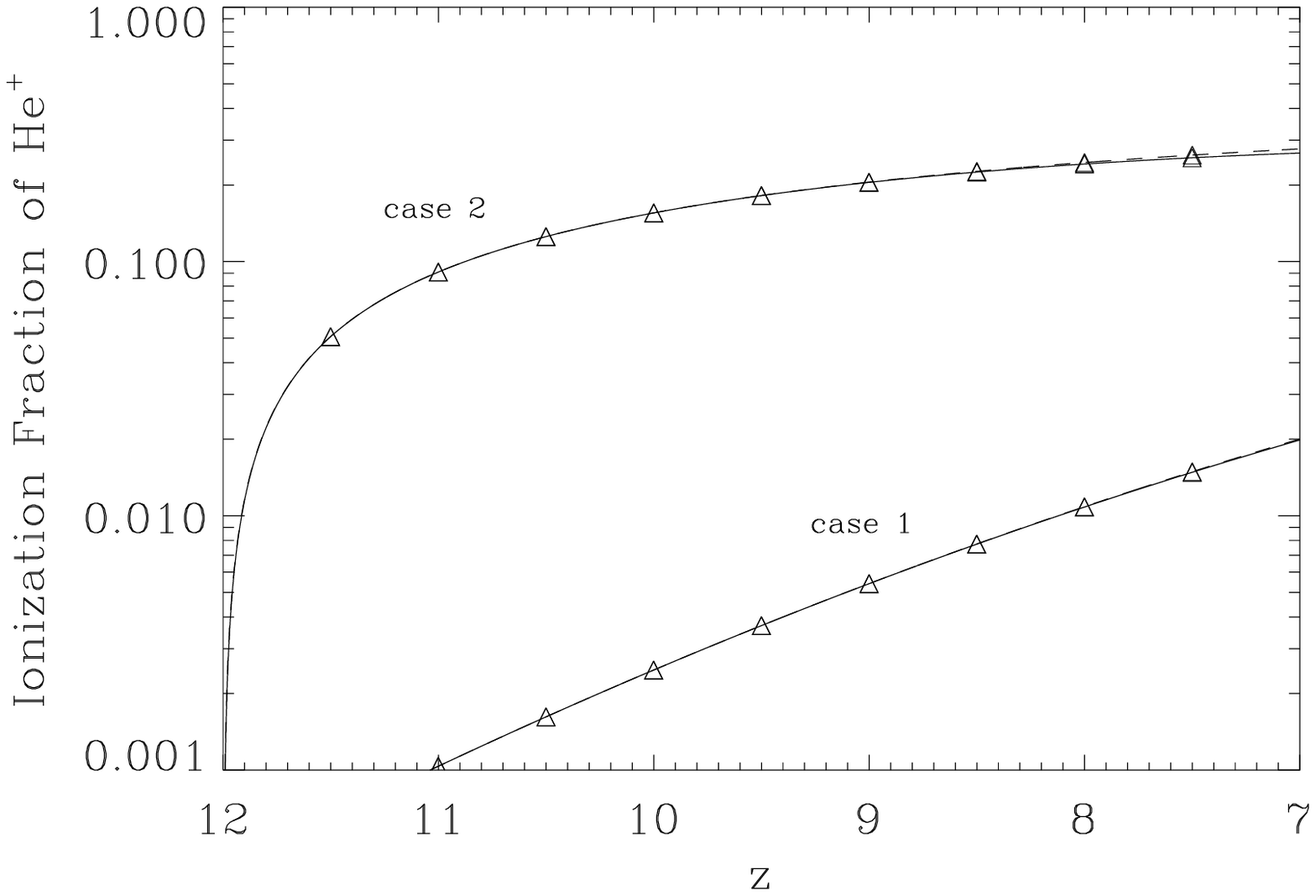}}}

\vspace{0.1in}
\figcaption{The IGM ionization fraction of He$^+$ is shown as a function
of redshift for the two XRBs described in the text, each with (dashed
lines) and without (solid lines) adiabatic cooling from cosmological
expansion.}
\vspace{0.2in}

We note here some consequences of our adopted photoionization cross
sections and of consistently including the secondary effects of
photoelectrons. The He~I cross-section is taken from the fit to
experimental data provided by \citet{verner}. This differs dramatically in
the keV energy range from the two-power-law fit from \citet{ost} used by
\citet{sk87} and \citet{ab97}, which is only valid for photon energies of a
factor of a few above the threshold value of 24.6 eV. This power-law
extrapolation exceeds the \citet{verner} He~I cross-section by a factor of
40 at 1 keV, and by a factor of 400 at 10 keV. It is therefore important to
use the latter fit for He~I when considering the effects of a pure
XRB. Furthermore, the ratio of the H~I/He~I photoionization cross sections
is very small in the X-ray range, about 3\% for photon energies of 1--10
keV. Thus, in the pre-reionization universe, when the IGM is predominantly
neutral and can be permeated only by X-rays, the ionization balance is
driven primarily by the photoionization of He~I and its associated
photoelectrons (and not those from H~I), each of which then ionize about a
dozen hydrogen atoms (SVS85). Since the primordial ratio of helium to
hydrogen is about 8\%, we may anticipate that $x_{\rm He^+} \sim x_{\rm
H^+}$ in such a scenario, and that is what we find.

For the purposes of the discussion below, we now take case 2 with
adiabatic cooling to be our fiducial case, as it provides an upper
bound to the magnitude of the results in Figures 1--3, or
equivalently, it represents the ``maximal effect'' of a high-$z$ XRB.
In this case, H~I ionization is dominated by secondary ionizations
from the He~I photoelectrons, as expected, followed by radiation from
the bound-bound transitions of He~I, while He~I is most affected by
direct photoionization and secondary ionizations from its own
photoelectrons. The major sources of IGM heating are, in decreasing
order, direct photoelectric heating of He~I, H~I and He~II, and
heating from the bound-bound transitions of He~I. At the earliest
epochs that we consider, adiabatic expansion and Compton cooling were
the dominant cooling processes.  For Case 2, by $z \sim 11$, the IGM
had experienced sufficient heating and ionization that line excitation
cooling was the most significant source of cooling.  For Case 1,
adiabatic expansion remained the dominant cooling term.

\subsection{Jeans Mass Filtering}

We now discuss the implications of such a heated, partially ionized IGM,
beginning with the evolution of overdensities in a warm IGM.  At any
redshift, dark matter perturbations with virial temperatures $T_{\rm
vir}(z)$ will decouple from universal expansion on the associated mass
scales and begin to collapse. The baryons will also fall into these dark
matter potential wells if they have little pressure associated with
them. If the gas in the IGM is heated by luminous sources, and if $T_{\rm
IGM}$ exceeds $T_{\rm vir}$, the baryons will resist collapse on the mass
scale set by $T_{\rm IGM}$, leading to a filtering of the mass scale that
can collapse and virialize. This feedback from a heated IGM has been
explored for a number of reheating scenarios
\citep{teg97,sgb94,valsilk}. As we do not account for gas clumping in our
analysis, we follow \citet{sgb94} in evaluating the magnitude of the
filtering for the cosmological Jeans mass.  Assuming that the temperature
of the baryons and the CMB are coupled until $z \sim 150$, and that the IGM
cools adiabatically after that, the Jeans mass in baryons is:

\begin{eqnarray}
M_{\rm J} \approx 2.2 \times 10^3 \, \left[\frac{\Omega_{\rm b}}{h \, (\mu \Omega_{\rm m})^{1.5}}\right] \, \left(\frac{1 + z}{10}\right)^{1.5} \, {\rm M}_\odot .
\end{eqnarray}
\vspace{0.05in}

If, however, the IGM is heated to a temperature $T_{\rm IGM}$, the Jeans
mass will scale as: 

\begin{eqnarray}
M_{\rm J} \approx 1.3 \times 10^5 \, \left[\frac{\Omega_{\rm b}}{h \, (\mu \Omega_{\rm m})^{1.5}}\right] \, \left(\frac{T_{\rm IGM}}{T_{\rm CMB}}\right)^{1.5} \, {\rm M}_\odot,
\end{eqnarray}
\vspace{0.05in}

\noindent where $T_{\rm CMB}$ = $T_{\rm CMB,0}$ $(1 + z)$. For the purposes
of comparison, we work with the fiducial case defined above.  At $z = 9$,
$T_{\rm IGM}$ $\sim$ 1.3 $\times 10^4$ K, so that $M_{\rm J}$ $\sim 3
\times 10^8$ $M_\odot$, as opposed to $M_{\rm J}$ $\sim$ 520 $M_\odot$ from
eqn. 4.  Thus, the Jeans mass in a pre-heated IGM exceeds by several orders
of magnitude its corresponding value in the case of an IGM that cools
adiabatically after decoupling from the CMB. Furthermore, the filtered
Jeans mass increases with decreasing $z$, unlike the canonical Jeans mass.

\subsection{The CMB}

A partially ionized IGM can affect the CMB through Thomson scattering
of the CMB photons off the free electrons in the IGM. There are at
least two potentially observable effects. First, the optical depth to
electron scattering, $\tau_{\rm e}$, as measured from the primary CMB
anisotropies, will have some contribution from the partially ionized
IGM preceding complete reionization, and may no longer be a strong
constraint on the epoch of reionization.  The expression for
$\tau_{\rm e}$ as a function of redshift in a $\Lambda$-cosmology is
given by (see, e.g., \citet{ts95}):

\begin{eqnarray}
\tau_{\rm e} (z) & \simeq &  0.057 \; \Omega_{\rm b} h \, \times \nonumber \\ & & \int_0^{z} dz^\prime \; \frac{(1 + z^\prime)^2 \, x_{\rm e}(z^\prime)}{\sqrt{\Omega_{\Lambda} + (1 + z^\prime)^2(1 - \Omega_{\Lambda} + \Omega_{\rm m} z^\prime)}} .
\end{eqnarray}
\vspace{0.05in}

\noindent This assumes that all the baryons are in a homogeneous IGM at all
times, which may not be a fair approximation at late epochs. As a simple
calculation, let us examine two possibilities: one, that reionization
occurred at $z = 6$, and that $x_{\rm H^+}$ is given by our standard XRB
for $7 \leq z \leq 12$, with a post-recombination residual ionization
fraction of 10$^{-4}$ for $12 < z < 1000$. The values of $x_{\rm H^+}$ are
linearly extrapolated between redshifts of 6 and 7. The second case assumes
that there is no high-$z$ XRB, and that $x_{\rm H^+}$ = 10$^{-4}$ for $7
\leq z \leq 1000$. Figure 4 displays $\tau_{\rm e}$ as a function of
redshift for these two scenarios. In the former case, the integrated
optical depth to recombination is $\tau_{\rm e}$ $\sim$ 0.047, of which
0.032, or about 68\%, is from the reionized IGM, and 0.009 comes from the
partially ionized IGM, roughly a 20\% effect. Though this seems a small
correction, it is a level of accuracy that can in principle be probed by
future CMB experiments such as $Planck$ \citep{eht99}. The additional
contribution from the partially ionized IGM could cause an overestimation
of the reionization epoch, $z_{\rm r}$, if the cumulative optical depth
were attributed solely to the IGM after complete reionization.
Unfortunately, the correction to $\tau_{\rm e}$ from this XRB has roughly
the same value as that from the IGM's residual electron fraction at $12 < z
< 1000$ ($\tau_{\rm e} \sim$ 0.006), due to the late reionization model and
a low total $\tau_{\rm e}$.  

\centerline{\epsfxsize=1.0\hsize{\epsfbox{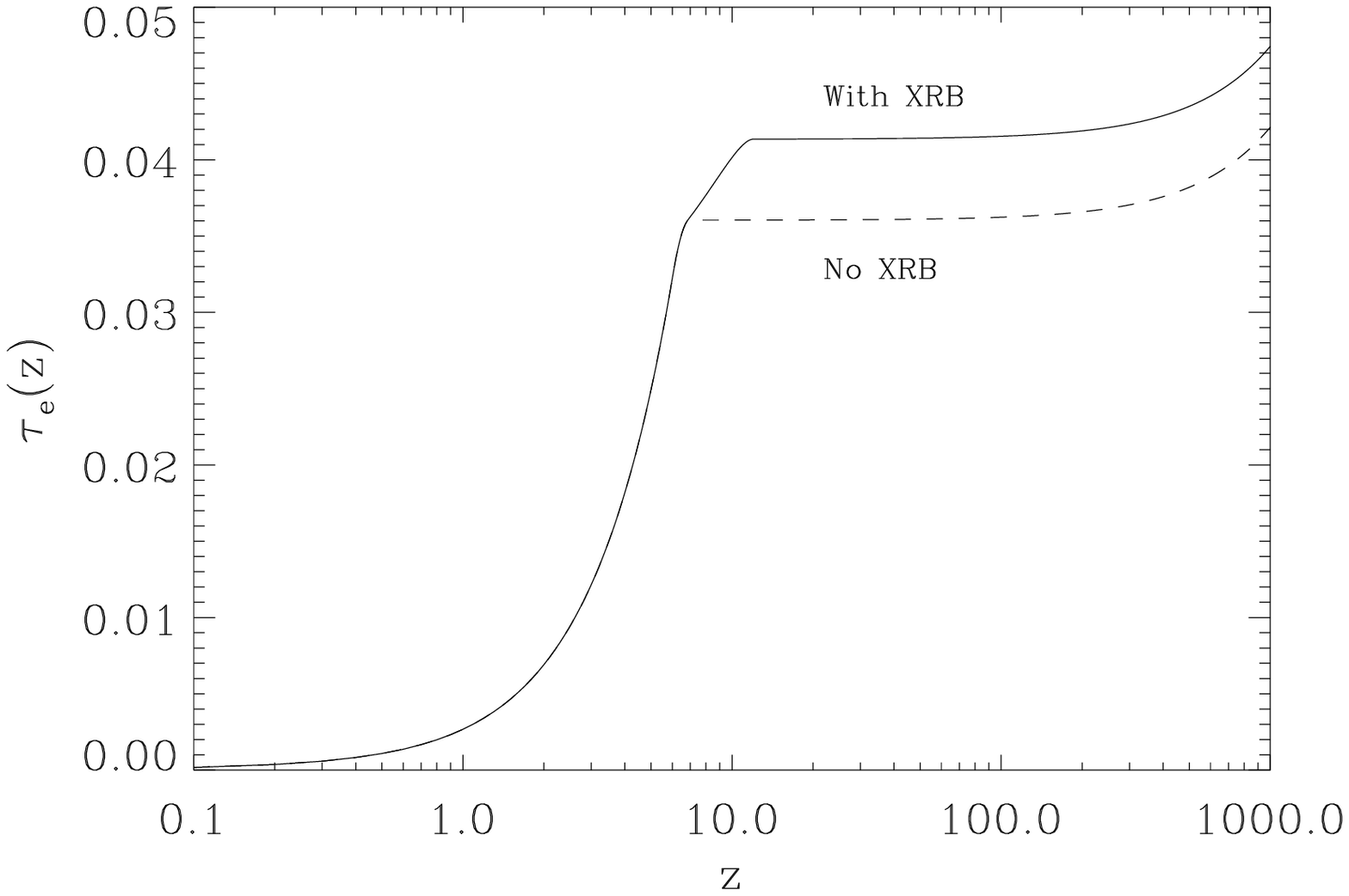}}}

\vspace{0.1in} 
\figcaption{The optical depth to electron scattering as a function of
redshift. Solid line assumes reionization of the IGM at $z = 6$,
partial ionization for $7 \leq z \leq 12$ for XRB case 2 with
adiabatic cooling, and the primordial electron freeze-in abundance at
higher redshifts. Dashed line represents the same scenario, but with
no XRB.}
\vspace{0.2in}

Since the theoretical uncertainty in the post-recombination $x_{\rm e}$ is
of order 10\% at present, all that is required to distinguish between the
``with XRB'' case in Figure 4 and a ``No XRB'' case with somewhat earlier
full reionization is an independent measurement of $z_{\rm r}$. While
$Planck$ could in principle map out the low-multipole ($l$) polarization
peak caused by reionization in the CMB angular power spectrum to
significant accuracy, for late reionization this feature is of sufficiently
low magnitude and at large enough scales ($l$ $\la$ 10) that foregrounds
are likely to be a complication. Ultimately, a spectroscopic determination
of the reionization epoch via the Gunn-Peterson effect from the SDSS or
from the Next Generation Space Telescope, in conjunction with future CMB
data, may provide the best probe of an IGM partially ionized by X-rays
prior to full reionization. The first of these observations is a real
possibility in the near future from efforts that are well underway with
data on SDSS QSOs at redshifts $\ga$ 5.8 \citep{fan3, becker}. Otherwise,
for all practical purposes, it may be a challenge to separate these two
different contributions to $\tau_{\rm e}$ from the epochs preceding
reionization.

Although a partially ionized IGM may smear a direct relation between
$\tau_{\rm e}$ and $z_{\rm r}$ as determined from the CMB, the majority of
CMB photons that are Thomson-scattered at late times are still scattered by
a fully ionized IGM. To see this, consider that the probability that a CMB
photon scattered off an IGM electron after redshift $z$ along the line of
sight is [1 - exp(-$\tau_{\rm e} (z)$)]. The derivative of this, called the
visibility function (see, e.g., \citet{tsb94}), is the relative probability
of the last scattering redshift of a CMB photon. We show the visibility
function in Figure 5, for the two scenarios in Figure 4. We see that,
despite the contribution to $\tau_{\rm e}$ from an IGM that is partially
ionized by X-rays, the surface of last scattering after recombination for
the CMB photons that we see today is still strongly biased towards the
epochs following complete reionization. Though we do not display it here,
we note that the peak in the visibility function corresponding to
recombination is much larger than that associated with reionization (see,
e.g., \citet{zal97b}), especially when the latter occurs at low redshifts.

\centerline{\epsfxsize=1.0\hsize{\epsfbox{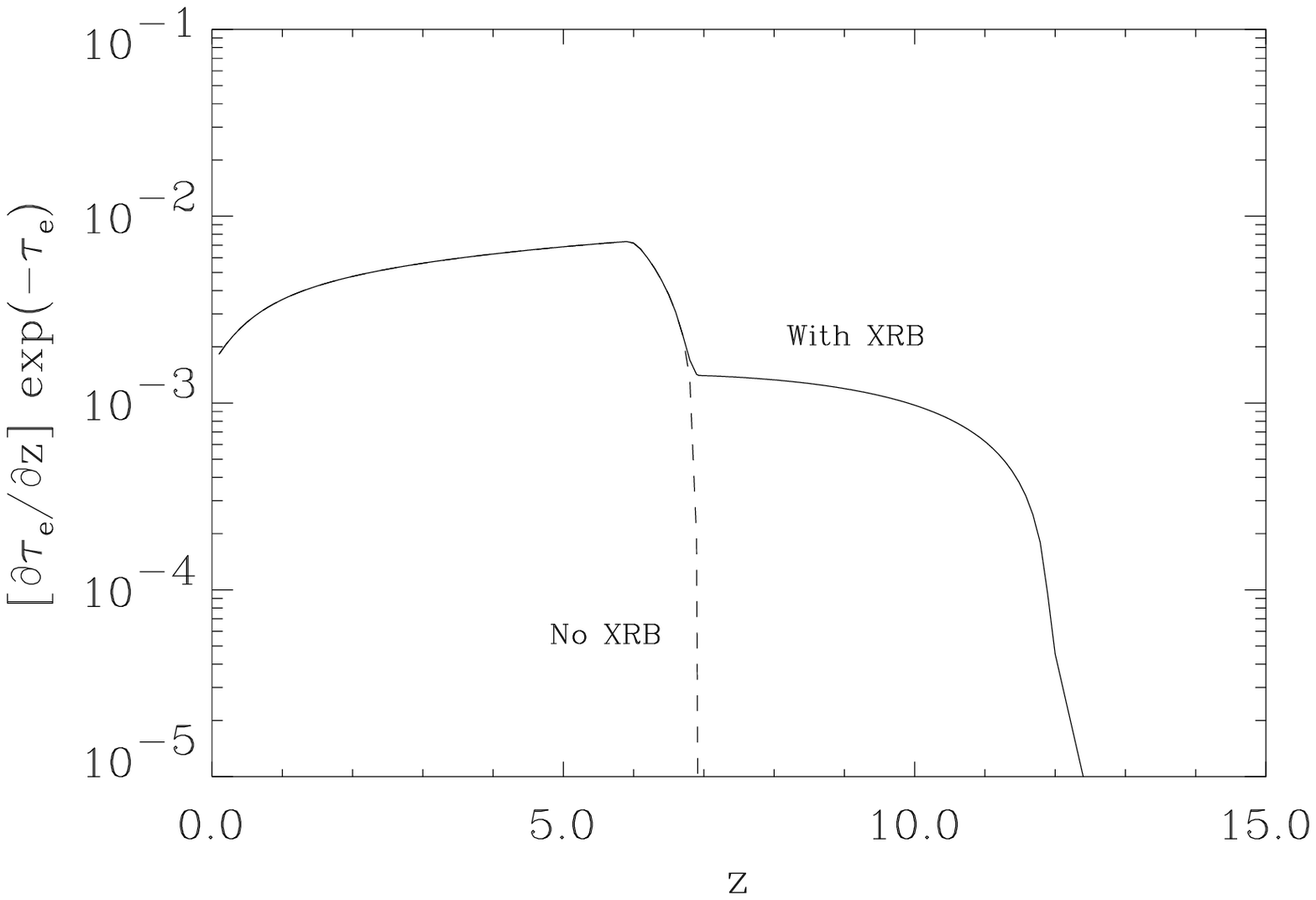}}}

\vspace{0.1in} 
\figcaption{The probability distribution of a CMB photon being
Thomson-scattered by an IGM electron as a function of redshift, with
the same assumptions as in Figure 4.}
\vspace{0.2in}

A second signature involves the correlations between ionized regions in a
patchy reionization model, which appear in the multipole moments of the CMB
anisotropies, typically at very small angular scales ($l \ga 5000$)
\citep{ksd98}. Since these correlations are usually predicted assuming a
neutral background IGM, one may pose the question of whether such
correlations may be significantly erased in the presence of a partially
ionized IGM.  As it happens, this is not a concern since the dominant
contribution to the above correlations come from patches of full
ionization, which are themselves highly correlated with high-density
regions. Thus, the relevant factor is the correlation of the underlying
matter density field, rather than the sizes of ionized ($x = 1$) regions or
the ionization contrast between $x = 1$ regions and the background IGM. The
relative contributions from these different effects is best seen in
\citet{valsilk2}.

\subsection{Molecular Hydrogen}

\begin{table*}[htb]
\renewcommand{\arraystretch}{1.2}
\begin{center}
\caption{Molecular Hydrogen Production}
\vspace{0.2in}
\begin{tabular}{rll}
\tableline \tableline
 Background & $n({\rm H}^-)/n_{\rm H}$ & $n({\rm H}_2)/n_{\rm H}$  \\ \tableline
XRB &  1.1 $\times 10^{-7}$ &  1.6 $\times 10^{-5}$ \\
XRB + IR/O &  3.3 $\times 10^{-12}$ &  4.8 $\times 10^{-10}$ \\
XRB + FUV &  1.1 $\times 10^{-7}$ &  4.2 $\times 10^{-10}$ \\
XRB + FUV + IR/O &  3.3 $\times 10^{-12}$ &  1.2 $\times 10^{-14}$ \\
\tableline
\end{tabular}

\tablecomments{The relative roles played by X-ray, FUV (Lyman-Werner
bands) and IR/optical backgrounds (see text for the individual definitions
of photon energy ranges) in an equilibrium calculation of n(H$^-$) and
n(H$_2$) for XRB case 2 with adiabatic cooling at $z = 9$.}
\end{center}
\end{table*}

We now turn to the implications of an X-ray heated, partially ionized IGM
for the production of molecular hydrogen, H$_2$, which is an important
coolant for primordial gas in virializing structures. A prerequisite for
this is that the baryons are able to collapse into virializing dark matter
halos, which may be hindered by a heated IGM, as discussed above. We
emphasize that we focus here on H$_2$ creation and destruction in the IGM,
rather than within dense collapsed structures. We include the following
processes in an equilibrium calculation of the abundance of H$_2$ relative
to hydrogen, in which we consider only the H$^{-}$ creation channel and do
not include formation via H$_2^+$ \citep{ls84,ab97}. Unless stated
otherwise, rates for individual reactions are taken from
\citet{donshull91}. H$^{-}$ is formed through radiative attachment of H~I
\citep{gp98}, and is destroyed by associative detachment by H~I, mutual
neutralization with protons, collisional detachment by electrons
\citep{ab97}, and photodetachment by photons of energy 0.755 to 13.6 eV. We
assume that H$_2$ forms by associative detachment of H~I and H$^{-}$, and
is destroyed by collisions with electrons and with H~I, by charge exchange
with protons, and by photodissociation in the Lyman and Werner bands by
photons of energies 11.2--13.6 eV.

As we do not specifically account for radiation below $\sim$ 1 keV, we
extrapolate the flux of each XRB at 2 keV down to 2500 \AA\ (4.97 eV)
using an effective X-ray spectral index of $\langle \alpha_{\rm 0,X}
\rangle$ = -- 1.38, with a $\nu^{-0.5}$ power-law for photon energies
below about 5 eV. We then compute the photodissociation rate of H$^-$
directly from the tabulated cross-sections of \citet{wishart} using
the fit given in \citet{teg97}, and of H$_2$ as described in
\citet{donshull91}, for the respective photon energy ranges stated
above. For both H$^{-}$ and H$_2$, particularly the latter, the
photodissociation rates exceed the other destruction channels by
several orders of magnitude. As we have performed an equilibrium
calculation and have not included the post-recombination freeze-in
abundance of H$_2$ relative to hydrogen of $\sim$ $10^{-6}$ in the
IGM, our results should be taken as representing the effect of an XRB
for newly created or destroyed H$_2$.

We find that, for XRB cases 1 and 2, n(H$_2$)/n$_{\rm H}$ never
exceeds about $3 \times 10^{-14}$, and has little evolution over the
redshifts that we consider; this falls well short of the critical
value of $\sim 10^{-6}$ to stimulate gas cooling in high-$z$
structures. Although the presence of X-rays boosts the free electron
fraction, at any given redshift the stronger XRB creates {\it less}
new H$_2$, owing to the combination of various dissociation processes
that are enhanced in a heated IGM, and to the strong role played by
photodissociation in the destruction of both H$^{-}$ and H$_2$. We
find that the timescales associated with the formation and
photodestruction of H$^{-}$ and H$_2$ are always much less than the
Hubble time at each epoch, so that our equilibrium calculations here
indicate real changes to the primordial freeze-in abundance of H$_2$.

To illustrate the relative roles played by an XRB and the associated
radiation fields below 13.6 eV, we display in Table 1 the equilibrium
abundances of H$^{-}$ and H$_2$ calculated by successively including
radiation from two backgrounds, defined as follows. The FUV includes the
photons relevant for H$_2$ photodissociation in the Lyman-Werner bands
(11.2--13.6 eV). The H$^{-}$ is dissociated by photons having energies of
0.755--13.6 eV, in the framework of this paper. For the purposes of Table
1, however, the IR/O (infrared/optical) term refers to the inclusion of all
sub-13.6 eV photons except those in the LW bands. Although this energy
range (0.755--11.2 eV) corresponds to some UV photons as well, most of the
photodissociation occurs at near IR and optical wavelengths, given the
nature of the H$^{-}$ cross section and our adopted power-law form for the
background intensity at these energies.

We see that the minimum IR/O and FUV backgrounds associated with any
putative XRB from high-$z$ quasars more than compensate for any positive
feedback from the XRB in the form of an increased electron fraction in the
IGM (see, however, \citet{rgshull01} on the positive feedback for H$_2$
formation in the vicinity of individual ionization fronts generated by hard
stellar spectra). The negative feedback from FUV radiation has already been
noted in several works; we point out here the additional feedback from IR/O
photons. The importance of the IR/O photons in this paper relative to the
results of previous works, e.g., \citet{har}, can be traced to at least two
factors. Firstly, our equilibrium calculation is performed for the
relatively low-density conditions in the IGM at $z = 9$, rather than within
highly overdense collapsed halos, so that some variance can be attributed
to the differing densities of the studied environments.  Secondly, Table 1
represents the effects of the strongest XRB considered here, which has a
correspondingly high IR/O associated background. An XRB that was
extrapolated from the observed EUV specific intensity at $z =3$, such as
case S in this work, would have a related IR/O background that has a
negligible impact on the net H$^{-}$, and hence H$_2$, abundances. Note
also that a combination of these two factors, IGM density and radiation
intensity at the H$^{-}$ photodissociation threshold, implies that only a
relatively modest IR/O background is required at $z < 12$ to destroy
H$^{-}$. A radiation field of far greater energy density is needed to
accomplish this in the early universe; this leads to the well-known result
that the CMB photons are believed to have photodissociated H$^{-}$ at $z
\ga 100$.

Our estimate of the FUV background is not necessarily the minimal value,
due to the opacity of the IGM in the Lyman-Werner bands which we have not
accounted for here. The table shows, however, that even an order of
magnitude reduction in the FUV background alone will not increase the
fractional abundance of H$_2$ above 10$^{-6}$. As we have neglected
contributions from stellar radiation, we have in fact underestimated the
effects of IR/O and FUV photons at the epochs we consider.  In summary, we
find that an XRB does not produce significant amounts of new H$_2$ in the
IGM, unless the associated FUV and IR/O backgrounds were somehow strongly
attenuated.

\subsection{Inhibition of Outflows}

Finally, it may be possible for a heated IGM with sufficient pressure to
inhibit outflows from star-forming protogalaxies. Mass loss from early
objects through galactic winds or evaporation is often invoked to explain
the ubiquitous presence of metals in the Ly$\alpha$ forest clouds at $z
\sim 3$; a pre-heated IGM could, however, hinder such outflows of
metal-enriched gas from host galaxies at sufficiently high redshifts (prior
to reionization). An estimate of the effect of a warm IGM in inhibiting
galactic winds may be made as follows.  Setting the ram pressure of the
wind $\rho_{\rm w} v_{\rm w}^2$, equal to $P_{\rm IGM}$ = ($n_{\rm H} +
n_{\rm He} + n_{\rm e}$) $k T_{\rm IGM}$, and using $\dot{M_{\rm w}} = 4
\pi r_{\rm st}^2 \, \rho_{\rm w} v_{\rm w}$, where $\dot{M_{\rm w}}$ and
$r_{\rm st}$ are respectively the mass outflow rate of the wind and the
wind stalling radius in the IGM where pressure equilibrium is reached, we
have:

\begin{eqnarray}
r_{\rm st} & \simeq & (0.5 \, {\rm Mpc}) \, \times \nonumber \\ & &  \left[ \left(\frac{\dot{M_{\rm w}}}{M_\odot/{\rm yr}}\right) \, \left(\frac{v_{\rm w}}{10^3 \, {\rm km/s}}\right) \, \left(\frac{10^4 {\rm K}}{T_{\rm IGM}}\right) \left(\frac{10}{1 + z}\right)^3 \, \right]^{1/2}
\end{eqnarray}
\vspace{0.05in}

\noindent for the IGM density at $z = 9$. Note that if the IGM had cooled
adiabatically from $z \sim 150$, then $T_{\rm IGM}$ would be a few Kelvin,
and $r_{\rm st}$ $\sim$ 33 Mpc. This neglects the added contribution that
any ionizing photons escaping from the protogalaxy would make to the local
ambient pressure.

Eqn. 7 approximately indicates the maximum degree, case 2 being the
strongest XRB considered here, to which an early XRB may hinder outflows
from starforming galaxies at $z = 9$. Note, however, that at lower
redshifts, e.g., $z \sim 3$, $r_{\rm st}$ can be significantly larger,
particularly for the mass outflow rates of a few to ten solar masses per
year that are typical of starbursting galaxies. Thus, the observations of a
trace metallicity in the IGM at $z \sim 3$ does not necessarily rule out an
early XRB from QSOs prior to reionization, if the epoch of metal ejection
associated with stellar activity in high-$z$ galaxies succeeds that of
pre-heating from an XRB by a sufficiently long period. This becomes a
particularly important constraint, however, for a model that posits an
early XRB from stars, or one that explores the heating effects of X-rays
for the post-reionization IGM \citep{me99}.

\section{Conclusions}

We have examined the effects of various high-$z$ XRBs, including a
case which mimics X-ray photons from early bright QSOs, on the
temperature and ionization of the IGM before reionization is
complete. We have found that individual luminous sources with their
associated H II/He III ionized regions may be embedded in a warm,
partially ionized IGM, rather than in a cold neutral IGM, during the
epochs before individual Str\"{o}mgren spheres have overlapped. 

The heating and ionizing effects from the XRBs were determined
self-consistently, including the associated generation of photoelectrons
which act as secondary sources, resulting in an enhanced population of free
electrons. We have investigated two XRB cases that are related to the
measured present-day soft XRB, and find that they can heat the IGM to
temperatures between a hundred to 10$^4$ K prior to reionization, and
result in ionization levels between fractions of a percent to about
20\%. The third XRB, assumed to be generated by early QSOs whose space
density evolves as indicated by the recent observations of bright high-$z$
quasars from the SDSS collaboration, did not appreciably alter the IGM
temperature and ionization from their post-recombination values at the
redshifts we considered.

We have examined the implications of our results for several phenomena of
interest to cosmology, including the Jeans mass, the CMB, and the
production of H$_2$. In particular, the pre-heated IGM raises the Jeans
mass significantly at the redshifts of interest to the growth of structure,
leading to a filtering of the baryonic mass scales that can collapse into
virializing dark matter halos. We also find that the XRB-enhanced electron
fraction in the IGM prior to reionization increases the total optical depth
to electron scattering. This could lead to an overestimation of the
reionization epoch if the cumulative $\tau_{\rm e}$ as measured by CMB
experiments were attributed solely to the fully ionized IGM.  However, the
addition to $\tau_{\rm e}$ from the post-recombination IGM, with a
freeze-in $x_{\rm e}$ $\sim 10^{-4}$ for $z \la 1000$, is roughly
comparable to that from an IGM which is partially ionized by our standard
XRB. It may be a challenge to observationally distinguish these disparate
contributions to $\tau_{\rm e}$ in late reionization scenarios, unless
there is an independent measurement of the reionization epoch,
e.g. spectroscopically through the Gunn-Peterson effect from the SDSS or
from future space telescopes.  Such a constraint seems particularly
promising from ongoing analyses that are well underway with SDSS data on
QSOs at $z \ga$ 5.8 \citep{fan3, becker}. In conjunction with forthcoming
CMB data, it could provide the best probe of an IGM partially ionized by
X-rays prior to full reionization, which otherwise may not leave a unique
imprint in the CMB. Finally, if the IGM is heated prior to reionization,
its thermal pressure could suppress outflows from galaxies. A pre-heated
IGM may also be detected by future radio telescopes in 21-cm emission
against the CMB \citep{tozzi}.

Despite the increased electron fraction in the IGM, the amount of
H$^{-}$-catalyzed molecular hydrogen formed is insignificant, contrary to
expectations. This is primarily due to the strong photodissociating effects
of the minimum IR/O and FUV (Lyman-Werner bands) backgrounds associated
with any high-$z$ XRB on H$^{-}$ and H$_2$ respectively. As we have
neglected radiation from stars in the treatment here, we have
underestimated these photodestruction terms at the redshifts we
consider. We therefore conclude that, unless the FUV and IR/O backgrounds
are strongly attenuated by some process which we have not included here
(see Table 1), the positive feedback from a boosted electron fraction
caused by a high-$z$ XRB is not sufficient to produce interesting amounts
of new H$_2$ in a uniform IGM.

We mention here that there are several issues that we have neglected or
oversimplified in this paper. First, we have not included the effects of
any clumping in the IGM, which will reduce our adiabatic cooling estimates,
alter the equilibrium calculation values of n(H$_2$), and challenge our
assumption that the X-rays heat the IGM homogeneously. Second, although we
have noted the Jeans mass filtering in a heated IGM, we have not examined
in detail the further cooling that must occur in virializing halos in order
for star formation to occur. We have found that an XRB may not create
significant amounts of new H$_2$ in the IGM, but the situation may prove to
be different within a halo or an IGM overdensity that is being irradiated
by X-rays from a nearby source (see, e.g., \citet{rgshull01}). In such
cases, the favored regimes of baryon density and temperature for H$_2$
production may change. It would be interesting to explore the feedback of a
heated partially ionized IGM on nonequilibrium H$_2$ chemistry in the
presence of a (destroying) FUV versus (enhancing) X-ray background photon
field \citep{har}, particularly when including the effects of secondary
electrons in a dense, neutral halo at high redshift. We defer more detailed
treatments of these problems and related consequences to future work.

\acknowledgements

We thank Phil Maloney for useful discussions, and the anonymous referee for
helpful comments. We gratefully acknowledge support from NASA LTSA grant
NAG5-7262.

\end{document}